\documentclass[preprint,jap,aip,amsmath,amssymb,showpacs,letterpaper,lengthcheck]{revtex4-1}
\usepackage[english]{babel}

\usepackage{graphicx}

\begin{document}

\title{Electron-phonon scattering dynamics in ferromagnets on ultrafast timescales: Influence of the phonon temperature}

\date{\today}

\author{Sven Essert}
\affiliation{Department of Physics and Research Center OPTIMAS, %
University of Kaiserslautern, P.O. Box 3094, 67653 Kaiserslautern, Germany}
\author{Hans Christian Schneider}
\email{hcsch@physik.uni-kl.de}
\affiliation{Department of Physics and Research Center OPTIMAS, %
University of Kaiserslautern, P.O. Box 3094, 67653 Kaiserslautern, Germany}
 
\begin{abstract}
The magnetization response of bulk ferromagnets after excitation by an ultrashort optical pulse is calculated using a dynamical model of the Elliott-Yafet type that includes the effects of the spin-orbit interaction in the ab-initio ferromagnetic band structure, the electron-phonon interaction at the level of Boltzmann scattering integrals, and dynamical changes in the temperature of the phonon bath. Using realistic parameters for the ultrashort optical pulse, the computed maximum magnetization quenching achievable with electron-phonon scattering in a fixed band structure is much smaller than the quenching observed in experiments. Heating of the phonon bath is found to not appreciably change the magnetization dynamics on ultrashort timescales.
\end{abstract}

\maketitle

\section{Introduction}

The magnetization dynamics of ferromagnets after ultrashort pulse excitation has been investigated experimentally and theoretically since its first observation 15 years ago.~\cite{Beaurepaire} Although a lot of progress has been made and a viable macroscopic model for the magnetization dynamics exists in the three-temperature model, there are still open questions regarding the underlying microscopic mechanism of the demagnetization as well as the details of the angular momentum transfer~\cite{Stamm} on ultrashort timescales. Additional interest in this topic is generated by its potential implications for laser-induced control of the magnetization and, possibly, all-optical switching.  Some promising microscopic explanations of magnetization dynamics involve mechanisms of the Elliott-Yafet type,~\cite{Steiauf2} in which the total spin polarization is reduced by scattering processes that change the average spin of the scattered electrons. In the presence of spin-orbit interaction this may be achieved by electron-phonon/impurity scattering,~\cite{Koopmans,Steiauf} as is usually assumed, but also by electron-electron,\cite{Krauss} electron-magnon,\cite{Carpene} scattering.

To clarify the importance of electron-phonon scattering to the ultrafast demagnetization of ferromagnets, we recently investigated this scattering mechanism in bulk ferromagnets. By solving the dynamical equations for the momentum-resolved carrier distributions we showed that the electron-phonon scattering dynamics in the fixed band structure cannot explain the observed magnetization dynamics for realistic pulse intensities.~\cite{prev} Apart from the fixed band structure, our earlier calculation used a bath assumption for the phonon system and kept the lattice temperature fixed at the value before the excitation. After excitation with appreciable laser intensities, however, the electrons should transfer energy to the lattice quickly, leading to a rise in lattice temperature. The heating of the lattice should, in turn, affect the electronic dynamics because the available phase space for transitions in the phonon system is changed. Here, we investigate this effect on the electronic dynamics by including the temperature of the phonon bath as an additional dynamical variable. As in Ref.~\onlinecite{prev}, the optical excitation and the electron-phonon scattering are described at the level of Boltzmann scattering integrals, and the band structure, electron-phonon interaction matrix elements, as well as dipole matrix elements are calculated from density-functional theory.

\section{Model}

A detailed account of our numerical method including the connection to ab-initio calculations can be found in Ref.~\onlinecite{prev}. In the following, we give a brief description of our theoretical approach, and show how a dynamical change of the phonon bath temperature is achieved in the framework of our model. We describe the ferromagnetic dynamics by momentum-resolved carrier distributions $n_{\vec{k}}^{\mu}(t)$ for all bands $\mu$ of interest, for which we integrate the dynamical equations
\begin{equation}
\frac{\partial}{\partial t}n^{\mu}_{\vec{k}} =
\frac{\partial}{\partial t}n^{\mu}_{\vec{k}}\bigr|_{\text{opt}}+
\frac{\partial}{\partial t}n^{\mu}_{\vec{k}}\bigr|_{\text{e-p}} 
\label{eq:dndt}
\end{equation}
with the following contributions from the optical excitation and electron-phonon scattering:
\begin{align}
\frac{\partial}{\partial t}n^{\mu}_{\vec{k}}\bigr|_\text{opt}= &|E(t)|^{2}\sum_{\nu}\mathcal{K}^{\text{opt}}_{\mu,\nu}(\vec{k})[n^{\nu}_{\vec{k}}-n^{\mu}_{\vec{k}}]
 \label{eq:dndt_opt}\\
\frac{\partial}{\partial t}n^{\mu}_{\vec{k}}\bigr|_\text{e-p}= &
 \sum_{\nu,\vec{k}'}\mathcal{K}^{\text{e-p}}_{\mu,\nu}(\vec{k},\vec{k'}) [1-n^{\mu}_{\vec{k}} ]n^{\nu}_{\vec{k}'}\nonumber \\
 &-\sum_{\nu,\vec{k}'}\mathcal{K}^{\text{e-p}}_{\nu,\mu}(\vec{k}',\vec{k}) n^{\mu}_{\vec{k}}[1-n^{\nu}_{\vec{k}'}] 
\label{eq:dndt_ep}
\end{align}
Here, $E(t)$ denotes the amplitude of the laser field. The scattering kernels $\mathcal{K}^\text{opt}$ and $\mathcal{K}^\text{e-p}$ contain all the occupation independent quantities for the optical and the scattering contribution, respectively.~\cite{prev} The scattering kernels $\mathcal{K}^{\text{e-p}}$ include the electron-phonon matrix elements, phonon and electron energy dispersions, and the phonon distributions. If one employs a bath approximation for the phonons and keeps the phonon temperature  $T_\text{p}$ fixed, these scattering kernels are \emph{independent} of time. In this paper, we still assume that the phonon system can be treated as a bath, but we characterize it by a time-dependent temperature $T_\text{p}=T_\text{p}(t)$. This leads to an implicit time dependence of the scattering kernel $\mathcal{K}^\text{e-p}$ in Eq.~\eqref{eq:dndt_ep} because it depends on the phonon temperature via the phonon occupation numbers $\tilde{n}_{\vec{q}}^{\lambda}=1/[\exp(\hbar\omega_{\vec{q}}^{\lambda}/k_{B}T_\text{p})-1]$.
Here, $k_{B}$ is Boltzmann's constant, and the phonon modes are labeled by their polarization $\lambda$ and wave vector $\vec{q}$. The computation of $\mathcal{K}^\text{e-p}(T_\text{p})$ for a given temperature of the phonon bath implies a sum over all allowed transitions weighted by the corresponding phonon occupation numbers. Instead of redoing this summation in Eq.~\eqref{eq:dndt_ep} in every numerical time step, which would vastly increase the CPU time demands, we note the smooth, monotonic dependence of $\tilde{n}_{\vec{q}}^{\lambda}$ on $T_\text{p}$, which suggests also a smooth dependence of their weighted sum $\mathcal{K}^\text{e-p}(T_\text{p})$. We therefore approximately replace $\mathcal{K}^\text{e-p}(T_\text{p})$ by a polynomial of degree $P$ which can be obtained from a least squares fit to values $\mathcal{K}^\text{e-p}(T_\text{p})$ precomputed for a ``grid'' of temperatures in the temperature interval of interest
$\mathcal{K}^\text{e-p}(T_\text{p})\approx \sum_{m=0}^P\mathcal{K}_m^\text{e-p}T_\text{p}^m$.

The dynamical equation of the temperature $T_\text{p}$ of the phonon bath is derived from the requirement of energy conservation, which holds for the system of electrons and phonons if one neglects losses to the environment. That is, a change in energy of the electronic system $E_\text{e}$ should lead to an equal, but opposite, change in the energy of the phonon system $E_\text{p}$,
\begin{equation}
\begin{split}
\frac{\partial}{\partial t}T_\text{p}&=\frac{1}{C_\text{p}(T_\text{p})}\frac{\partial E_\text{p}}{\partial t}
=-\frac{1}{C_\text{p}(T_\text{p})}\left.\frac{\partial E_\text{e}}{\partial t}\right|_\text{e-p}\\
&=-\frac{1}{C_\text{p}(T_\text{p})}\sum_{\mu,\vec{k}} \epsilon^{\mu}_{\vec{k}} \frac{\partial}{\partial t} n^{\mu}_{\vec{k}} \Bigr|_\text{e-p},
\label{eq:dTdt}
\end{split}
\end{equation}
where $\epsilon^{\mu}_{\vec{k}}$ is the energy of single-particle state $|\mu,\vec{k}\rangle $ and $(\partial/\partial t)  n^{\mu}_{\vec{k}}|_\text{e-p}$ the scattering contribution to the occupation change as in Eq.~\eqref{eq:dndt_ep}. Finally, the phonon heat capacity
\begin{equation}
C_\text{p}(T_\text{p})=\frac{\partial E_\text{p}(T_\text{p})}{\partial T_\text{p}}=\sum_{\vec{q},\lambda}\hbar\omega_{\vec{q}}^{\lambda}\frac{\partial \tilde{n}_{\vec{q}}^{\lambda}(T_\text{p})}{\partial T_\text{p}}
\end{equation}
is computed using the phonon frequencies $\omega_{\vec{q}}^{\lambda}$ obtained from an ab-initio calculation.~\cite{prev} As $C_\text{p}(T_\text{p})$ also smoothly depends on the temperature $T_\text{p}$, it is  replaced in the numerical calculations by an polynomial approximation for the same set of ``grid'' points as $\mathcal{K}^\text{e-p}(T_\text{p})$. The combined set of dynamical equations for the electronic distributions, Eq.~\eqref{eq:dndt}, and the temperature $T_\text{p}$ of the phonon bath, Eq~\eqref{eq:dTdt}, is solved numerically. From the solution, the magnetization of the system can be obtained by the sum
\begin{equation}
 M(t)=\frac{2\mu_B}{\hbar}\sum_{\mu,\vec{k}} \langle S_z\rangle_{\mu,\vec{k}} \,n^{\mu}_{\vec{k}}(t),
\end{equation}
over the spin expectation values  $\langle S\rangle_{\mu,\vec{k}}$ of the states~$|\mu,\vec{k}\rangle$, neglecting orbital contributions to the magnetization. Here $\mu_B$ denotes the Bohr magneton.

\section{Results}

We apply the model and calculation procedure described in the previous section to the simple ferromagnets nickel and iron, and compare with our earlier results~\cite{prev} that assumed a constant temperature of the phonon bath. The parameters were chosen as in Ref.~\onlinecite{prev}. For the polynomial interpolations, we evaluated $\mathcal{K}^\text{e-p}$ and $C_\text{p}$ for 40 equidistant temperatures in the range from 250 to 1000\,K, and used an interpolation with third and fifth degree polynomials, respectively.

The results shown in Figs.~1 and~2 for nickel and iron, respectively, are computed assuming the same pulse intensity of 4 mJ/cm$^{2}$, from which the amplitude of the electric field in the material is obtained by material specific values for the complex refractive index. The optical excitation leads to a deposition of energy in the material that is quite similar for both nickel and iron, cf.~Figs.~1(b) and 2(b). The experimentally observed differences between nickel~\cite{Krauss} and iron~\cite{Carpene} at the same pump fluence are therefore not simply due to a different ``net'' energy transfer to the ferromagnet by the ultrashort pulse. The magnetization dynamics exhibits a rapid drop with different time constants for iron and nickel of less than 250\,fs, and a subsequent return to an equilibrium state on a time scale of 10\,ps. The demagnetization occurs via the so-called the Elliott-Yafet mechanism: The spin mixing due to the spin-orbit interaction leads to a change of the carriers' average spin in each scattering event and thus to a change of magnetization. On the longer time-scale the carrier distributions approach again an equilibrium Fermi-Dirac form and thus lead to a return of the magnetization towards an equilibrium value. 

As stressed in Ref.~\onlinecite{prev}, the computed magnetization dynamics should be compared to experimental results~\cite{Koopmans,Carpene,Krauss} on a sub-picosecond timescale, because changes in the quasi-equilibrium magnetization (exchange splitting), which dominate the dynamics on a picosecond timescale, see, e.g., Ref.~\onlinecite{Djordjevic}, are excluded from our model due to the fixed band structure. We also neglect carrier-carrier scattering~\cite{Krauss} which leads to an additional relaxation towards quasi-equilibrium distributions. The combination of ``magnetic'' dynamics and electron-electron scattering likely prevents the occurrence of  a dynamical magnetization larger than its equilibrium value, as it is the case for nickel around 4\,ps in our simulation. Figs. 1(a) and 2(a) indicate that the dynamical changes in the temperature of the phonon bath have no pronounced influence on the ultrafast magnetization dynamics. 
Still, the calculation including a time-dependent temperature of the phonon bath yields a slightly accelerated magnetization dynamics on the ultrafast timescale because the heating of the phonon bath increases the scattering rates. 

Our calculated results on a timescale of a few picoseconds should be interpreted strictly as a model study. Figs.~1(b),(c) and 2(b),(c) show that
the ``heating dynamics'' of nickel and iron are very similar, even though the magnetization dynamics are distinctly different. In both materials,  the system reaches a steady state on a timescale of several (ten) picoseconds when the temperature of the phonon bath is allowed to change dynamically. This steady state is characterized by a heated phonon system and an higher energy (density) of the electron system, because system does not reach the initial equilibrium distribution at $T_0=300\,\text{K}$, but instead one at the higher final temperature, which corresponds to a higher energy (density).  For nickel, as shown in Fig.~1, the dynamical phonon temperature even rises above the Curie temperature ($T_C(\text{Ni})=631\,\text{K}$)\cite{Gray1972}.  The deposited energy of about $130\,\text{meV} / \text{cell}$ is therefore enough to completely demagnetize the material after thermalization. Such a full demagnetization is not observed in experiment at the given intensity, which indicates that the constant amplitude of the optical field assumed by us is an overestimation. In the experiment the field is  attenuated due to absorption in the sample.  

\begin{figure}[h]
\includegraphics[width=9cm]{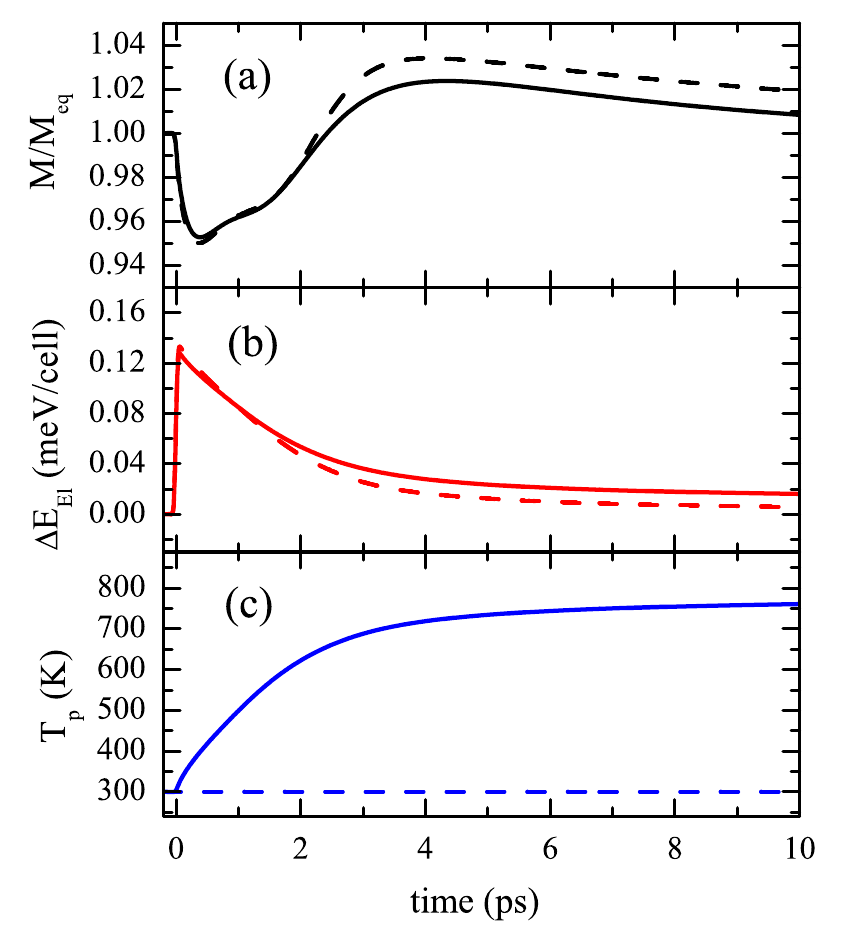}

\caption{\label{fig:ni}(a) Magnetization dynamics after optical excitation for \emph{nickel} computed with a dynamical (solid) and a fixed (dashed) temperature of the phonon bath; (b) energy in the electronic system, and (c) temperature of the phonon system.}
\end{figure}
\begin{figure}
\includegraphics[width=9cm]{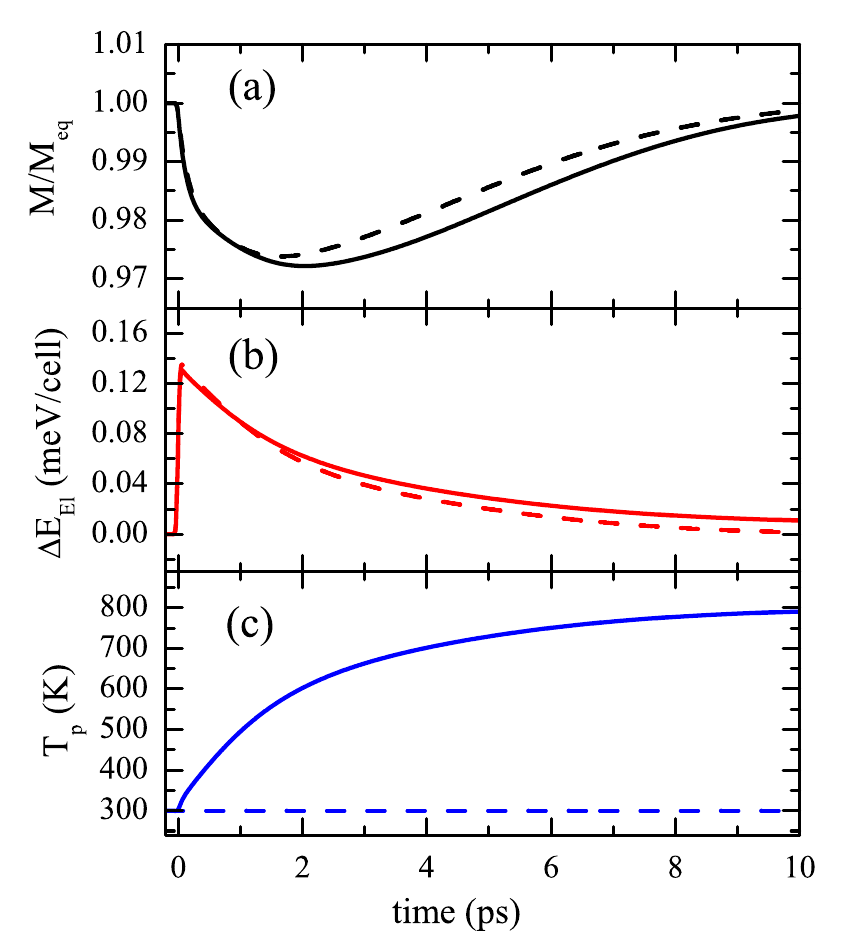}

\caption{\label{fig:fe}Magnetization dynamics after optical excitation for \emph{iron} computed with a dynamical (solid) and a fixed (dashed) temperature of the phonon bath; (b) energy in the electronic system, and (c) temperature of the phonon system.}
\end{figure}

\section{Conclusions}
We presented an ab-initio based calculation of electron-phonon scattering dynamics with a dynamical temperature of the phonon bath for nickel and iron, which constitutes a more realistic description of a thin film weakly coupled to a substrate than the fixed temperature of the phonon bath used in a previous study.~\cite{prev} In the framework of this model, which uses a fixed band structure, we compared the resulting demagnetization scenario with constant and time-dependent temperature of the phonon bath to estimate the influence of the phonon temperature on the dynamics. We find only marginal changes in the magnetization dynamics. This supports our earlier conclusion~\cite{prev} that electron-phonon scattering in a constant band structure is not able to explain the pronounced observed demagnetization for realistic laser intensities. Our results provide a hint that a dynamical phonon-bath temperature is unlikely to play an important role in the ultrafast magnetization dynamics of ferromagnets. Based on our results, we conclude that a dynamical change of the band structure, i.e., the exchange splitting has to be taken into account to achieve a quantitative understanding of ultrafast demagnetization dynamics in ferromagnets.

\end{document}